%% file: main.tex
\PassOptionsToPackage{unicode}{hyperref}
\PassOptionsToPackage{hyphens}{url}
\PassOptionsToPackage{dvipsnames,svgnames,x11names}{xcolor}
\documentclass[
  12pt]{article}

\input{macros}

\newcommand{\anon}{1}

\begin{document}


\if1\anon
{
  \title{\bf A Design-based Solution for Causal Inference with Text: Can a Language Model Be Too Large?}
  \author{Graham Tierney\thanks{
    GT and SK are joint first authors. GT conducted this work while at Duke University. The work was partially supported by a grant from the Templeton Foundation and a NSF CAREER award. Code and data for reproducibility can be found at \href{https://github.com/kattasa/text_as_treatment_repository/tree/main}{https://github.com/kattasa/text\_as\_treatment\_repository/tree/main}.}\hspace{.2cm}\\
    Netflix\\
    Srikar Katta$^*$ \\
    Department of Computer Science, Duke University\\
    Christopher Bail\\
    Department of Sociology, Duke University\\
    Sunshine Hillygus\\
    Department of Political Science, Duke University\\
    Alexander Volfovsky\\
    Dpeartment of Statistical Science, Duke University}
  \maketitle
} \fi

\if0\anon
{
  \bigskip
  \bigskip
  \bigskip
  \begin{center}
    {\LARGE\bf A Design-based Solution for Causal Inference with Text: Can a Language Model Be Too Large?}
\end{center}
  \medskip
} \fi

\bigskip
\begin{abstract}
Many social science questions ask how linguistic properties causally affect an audience’s attitudes and behaviors. Because text properties are often interlinked (e.g., angry reviews use profane language), we must control for possible latent confounding to isolate causal effects. Recent literature proposes adapting large language models (LLMs) to learn latent representations of text that successfully predict both treatment and the outcome. However, because the treatment is a component of the text, these deep learning methods risk learning representations that actually encode the treatment itself, inducing overlap bias. Rather than depending on post-hoc adjustments, we introduce a new experimental design that handles latent confounding, avoids the overlap issue, and unbiasedly estimates treatment effects. We apply this design in an experiment evaluating the persuasiveness of expressing humility in political communication. Methodologically, we demonstrate that LLM-based methods perform worse than even simple bag-of-words models using our real text and outcomes from our experiment. Substantively, we isolate the causal effect of expressing humility on the perceived persuasiveness of political statements, offering new insights on communication effects for social media platforms, policy makers, and social scientists.
\if0\anon{
\footnote{Code and data for reproducibility can be found at \href{https://github.com/qwertyanon/text_as_treatment_anon}{\texttt{https://github.com/qwertyanon/text\_as\_treatment\_anon}}.}
}\fi
\end{abstract}

\doublespacing
\section{Introduction}
\input{introduction}

\subsection{Operationalizing Linguistic Properties} \label{sec:ih_outcomes}
\input{survey_details} 

\section{Benchmarking Text-as-Treatment Estimators} \label{sec:sims}
\input{simulations}

\section{Experimental Design} \label{sec:experimental_design}
\input{experimental_design}

\section{Detailed Analysis} \label{sec:detailed_analysis}
\input{detailed_analysis}

\section{Conclusion}
\input{conclusion}

\bibliographystyle{agsm}
\bibliography{references}

\appendix
\section{Appendix}
\input{appendix}

\end{document}

%% file: macros.tex
\usepackage{iftex}
\ifPDFTeX
  \usepackage[T1]{fontenc}
  \usepackage[utf8]{inputenc}
  \usepackage{textcomp} 
\else 
  \usepackage{unicode-math}
  \defaultfontfeatures{Scale=MatchLowercase}
  \defaultfontfeatures[\rmfamily]{Ligatures=TeX,Scale=1}
\fi
\usepackage{lmodern}
\ifPDFTeX\else  
\fi
\IfFileExists{upquote.sty}{\usepackage{upquote}}{}
\IfFileExists{microtype.sty}{
  \usepackage[]{microtype}
  \UseMicrotypeSet[protrusion]{basicmath} 
}{}
\makeatletter
\@ifundefined{KOMAClassName}{
  \IfFileExists{parskip.sty}{%
    \usepackage{parskip}
  }{
    \setlength{\parindent}{0pt}
    \setlength{\parskip}{6pt plus 2pt minus 1pt}}
}{
  \KOMAoptions{parskip=half}}
\makeatother
\usepackage{xcolor}
\makeatletter
\ifx\paragraph\undefined\else
  \let\oldparagraph\paragraph
  \renewcommand{\paragraph}{
    \@ifstar
      \xxxParagraphStar
      \xxxParagraphNoStar
  }
  \newcommand{\xxxParagraphStar}[1]{\oldparagraph*{#1}\mbox{}}
  \newcommand{\xxxParagraphNoStar}[1]{\oldparagraph{#1}\mbox{}}
\fi
\ifx\subparagraph\undefined\else
  \let\oldsubparagraph\subparagraph
  \renewcommand{\subparagraph}{
    \@ifstar
      \xxxSubParagraphStar
      \xxxSubParagraphNoStar
  }
  \newcommand{\xxxSubParagraphStar}[1]{\oldsubparagraph*{#1}\mbox{}}
  \newcommand{\xxxSubParagraphNoStar}[1]{\oldsubparagraph{#1}\mbox{}}
\fi
\makeatother

\usepackage{longtable,booktabs,array}
\usepackage{calc} 
\usepackage{etoolbox}
\makeatletter
\patchcmd\longtable{\par}{\if@noskipsec\mbox{}\fi\par}{}{}
\makeatother
\IfFileExists{footnotehyper.sty}{\usepackage{footnotehyper}}{\usepackage{footnote}}
\makesavenoteenv{longtable}
\usepackage{graphicx}
\makeatletter
\def\maxwidth{\ifdim\Gin@nat@width>\linewidth\linewidth\else\Gin@nat@width\fi}
\def\maxheight{\ifdim\Gin@nat@height>\textheight\textheight\else\Gin@nat@height\fi}
\makeatother
\setkeys{Gin}{width=\maxwidth,height=\maxheight,keepaspectratio}
\makeatletter
\def\fps@figure{htbp}
\makeatother

\makeatletter
\@ifpackageloaded{caption}{}{\usepackage{caption}}
\AtBeginDocument{%
\ifdefined\contentsname
  \renewcommand*\contentsname{Table of contents}
\else
  \newcommand\contentsname{Table of contents}
\fi
\ifdefined\listfigurename
  \renewcommand*\listfigurename{List of Figures}
\else
  \newcommand\listfigurename{List of Figures}
\fi
\ifdefined\listtablename
  \renewcommand*\listtablename{List of Tables}
\else
  \newcommand\listtablename{List of Tables}
\fi
\ifdefined\figurename
  \renewcommand*\figurename{Figure}
\else
  \newcommand\figurename{Figure}
\fi
\ifdefined\tablename
  \renewcommand*\tablename{Table}
\else
  \newcommand\tablename{Table}
\fi
}
\@ifpackageloaded{float}{}{\usepackage{float}}
\floatstyle{ruled}
\@ifundefined{c@chapter}{\newfloat{codelisting}{h}{lop}}{\newfloat{codelisting}{h}{lop}[chapter]}
\floatname{codelisting}{Listing}

\makeatother
\makeatletter
\@ifpackageloaded{caption}{}{\usepackage{caption}}
\@ifpackageloaded{subcaption}{}{\usepackage{subcaption}}
\makeatother

\ifLuaTeX
  \usepackage{selnolig}  
\fi
\usepackage{bookmark}

\IfFileExists{xurl.sty}{\usepackage{xurl}}{} 
\hypersetup{
  pdftitle={Title},
  pdfauthor={Author 1; Author 2},
  pdfkeywords={3 to 6 keywords, that do not appear in the title},
  colorlinks=true,
  linkcolor={blue},
  filecolor={Maroon},
  citecolor={Blue},
  urlcolor={Blue},
  pdfcreator={LaTeX via pandoc}}

\usepackage{enumerate}
\usepackage{amsfonts, amsmath, amssymb, amsthm}
\usepackage{tabularx}
\usepackage{dcolumn}

\usepackage{natbib}

\usepackage{psfrag,epsf}

\usepackage{tikz}
\usetikzlibrary{bayesnet}
\usepackage[english]{babel}

\newtheorem{proposition}{Proposition}[section]

\usepackage{setspace}
\usepackage{appendix}

\newcommand{\arrogant}[1]{\underline{#1}}
\usepackage[normalem]{ulem}
\newcommand{\humble}[1]{\dotuline{#1}}
\newcommand{\colorful}[1]{\fbox{#1}}

\usepackage{enumitem}
\setlist[itemize]{topsep=1pt, itemsep=1pt}


%% file: introduction.tex
Many of the most pressing research topics about society and politics are fundamentally about communication. Political dialogue is considered a cornerstone of a healthy democracy, but today's political environment is instead characterized by polarization, overconfidence, and dogma \citep{dellaposta2015liberals, finkel2020political, iyengar2019origins, rathje2021out}. Designing solutions to bridge the political divide and to encourage more productive language requires a rigorous understanding of how linguistic properties can \textit{causally affect} behavior.

Among the many linguistic features that may influence political discourse, intellectual humility (IH) -- the capacity of people to recognize the fallibility of their beliefs and recognize the intellectual strengths of others -- has emerged as a promising candidate for promoting more constructive dialogue. We therefore center our subsequent discussion on examining the causal impact of expressing intellectual humility in political arguments on reader's perceptions of the arguments.
However, characterizing the behavioral effects of expressing IH is highly complex, as linguistic properties are often deeply intertwined -- word choice, intonation, punctuation, and other linguistic features all simultaneously contribute to perceptions of texts. Arguments that vary in their expression of IH typically differ along multiple linguistic dimensions, complicating subsequent causal inference.

To illustrate why the interconnectedness between linguistic properties is problematic, consider the open ended survey responses about gun control exhibiting differing levels of intellectual humility displayed in Table~\ref{tab:text_examples}. The response in the first column is both not intellectually humble \textit{and} uses descriptive language, like ``weapons of murder,'' to try and convince the reader to support gun control. In contrast, the second text in Table~\ref{tab:text_examples} does not employ such descriptive language and is intellectually humble. Existing work in communication -- especially intellectual humility in particular --  randomly assign some subjects to read the arrogant text and others to read the humble text and measure the readers' perceptions \citep{stroud2025intellectual}. Because the texts differ along multiple axes, it is unclear whether any difference in perceptions would be driven by expressing IH or by employing colorful language because the texts differ along both axes.

In other words, latent linguistic properties \textit{confound} the relationship between expressing IH and perceptions of the text.

\begin{table}[t]
\centering
\caption{Real arguments from participants about passing gun control legislation. The left text is both \arrogant{intellectually arrogant} and uses \colorful{colorful language}  while the \humble{intellectually humble} argument does not use any colorful language. Therefore, it is unclear if differences in perceptions of these arguments is due to humility or colorfulness. We have a latent confounding issue. Further, \humble{humility} and \arrogant{arrogance} are both text components; a language model that aims to encode text components predictive of treatment in its latent space will encode the treatment, thereby inducing an overlap bias.} \label{tab:text_examples}
\begin{tabularx}{\textwidth}{X|X}
\hline
\arrogant{Intellectually arrogant argument} & \humble{Intellectually humble argument} \\
\hline
Gun control \arrogant{needs to be} passed immediately. \arrogant{I know} that guns are \colorful{weapons of murder} that need to be controlled. Guns \arrogant{do not} protect people.& 
Gun control is a controversial topic. I'm \humble{not sure} how to best address the issue, but passing some basic gun control legislation \humble{could be} a start. \color{black}\\
\hline
\end{tabularx}
\end{table}

Several recent papers attempt to overcome this confounding issue by developing methods that, under certain additional assumptions, can identify causal effects with latent treatments \citep{pryzant2020causal, gui2022causal, fong2023causal, imai2024causal}. Broadly, these methods fall into two categories: (1) covariate adjustment approaches, which attempt to \textit{learn} confounders through some text representation, and (2) careful curation of texts to ensure that latent confounders are independent of treatment.

\paragraph*{Covariate adjustment methods} These approaches use language models to map the text into a lower-dimensional space that can be used as balancing scores \citep{assaad2021counterfactual}. However, covariate adjustment faces two competing challenges. First, if the language model fails to capture all relevant confounders, the resulting adjustments will be biased. For example, bag-of-words or topic model representations do not capture subtle but important linguistic properties like tone or context \citep{mozer2020matching, mozer2023leveraging, keith2020text}. In contrast, large language models (LLMs), which can capture more complex textual relationships, offer a promising path toward overcoming omitted variable bias \citep{veitch2020adapting, gui2022causal, pryzant2020causal, feder2022causal, lin2023text, imai2024causal}.

The second challenge concerns the \textit{overlap} assumption: ensuring that treated and control units are sufficiently comparable in the lower-dimensional space permits valid counterfactual estimation without extrapolation. While LLM-based approaches may successfully overcome omitted variable bias, their representation learning objectives may force them to induce overlap violations. For instance, \citet{gui2022causal} and \citet{veitch2020adapting} learn representations of the text that are predictive of both the treatment and the outcome. Because the treatment is a text component, these approaches run the risk of embedding the treatment itself in the text representations. In this case, LLM-based approaches inadvertently induce overlap biases. We demonstrate through simulations that language models encounter precisely these overlap challenges.

\paragraph*{Curation approaches} Avoiding the use of black box language models, \citet{fong2023causal} introduce an experimental design curating an existing corpus of texts. Here, the researcher will specify possible latent confounders and find texts in the corpus that are similar along all specified confounders \textit{except for} treatment. The researcher can then randomly assign readers to read either the treated or control texts. However, such an approach assumes the existence of an existing corpus and requires knowledge about latent confounding, which are not always feasible.

Additionally, \citet{imai2024causal} considers an experimental design that uses LLMs to generate texts that are similar on all features except for the treatment. While this approach presents an interesting opportunity at the interface of LLMs and experimental design, identification rests on a key assumption that the dimensionality of latent confounding in the LLM's representation is smaller than the dimensionality of latent treatment. Because LLMs are black boxes, such an assumption is untestable.

\paragraph*{Our approach} 
Avoiding the use of black box language models and the need for a pre-existing corpus of texts, we introduce a new experimental design that allows scientists to handle both omitted variable bias \textit{and} the overlap issue. Our approach uses study/survey participants rather than an existing corpus to produce natural texts that vary on a treatment feature; another set of study participants then edit these messages to change the treatment while holding all else fixed. This procedure gives us much wider and more natural variation in the expression of the latent treatment and other confounding features. And because this procedure leverages study participants to generate and edit texts, our approach is also entirely auditable: an auditor could look at both the original and edited text and judge their comparability to ensure we are making apples-to-apples comparisons.
We operationalize our experimental design to collect 6,994 evaluations of 1,830 unique texts to study the effect of expressing intellectual humility on perceptions of political arguments related to climate change, gun control, and immigration. Our results point to an apparent paradox: while humble language might soften perceptions of aggression--potentially fostering a more approachable and civil dialogue--it also potentially dilutes the effectiveness of the communication by reducing the perceived informativeness and persuasiveness, suggesting that intellectual humility may not be the best strategy for combating political polarization and dogmatism.

Our paper is organized in a slightly unorthodox way. We first begin by defining intellectual humility and explaining our survey outcomes in Section~\ref{sec:ih_outcomes}; the reader interested only in the methodological contribution can skip this section. We then benchmark current text-as-treatment estimators by designing simulations using our experimental data; in doing so, we ensure that we have a trustworthy ground-truth for the average treatment effect and realistic texts. Following the simulations, Section~\ref{sec:experimental_design} discusses identification, a new experimental design, and estimation that handles both latent confounding and the overlap issue. We return to the details of the experiment we performed in Section~\ref{sec:detailed_analysis}, where we share new discoveries on the role of intellectual humility in the effectiveness of political arguments.

%% file: survey_details.tex
Given the central role of intellectual humility in our analysis, we begin by defining this concept and explaining how we measure it in our survey outcomes. Intellectual humility (IH), the capacity of people to recognize the fallibility of their beliefs and recognize the intellectual strengths of others, involves (1) showing respect for others' beliefs and ideas, (2) considering counterpoints to one's own views, and (3) admitting limitations and uncertainties in one's own beliefs.

There is an emerging literature that identifies a correlation between negative political behavior and intellectual humility. For example, individuals low in intellectual humility are more likely to demonize political opponents and self-segregate into ideologically homogeneous social networks \citep{stanley2020intellectual}. Those with high intellectual humility, on the other hand, are more likely to connect with others who do not share their views \citep{stanley2020intellectual}. While these studies conceptualize IH as a \textit{personality trait}, more recent work has begun to examine its role as a linguistic property. For example, \citet{stroud2025intellectual} study the effects of expressing IH in political messages on affective polarization. However, their experimental design does not address latent confounding, limiting its ability to support \textit{causal} claims about the effects of expressing IH.

To this aim, we conducted a survey experiment studying the effects of expressing intellectual humility on audiences' perceptions of political arguments. We provided the survey reviewers a political argument and asked them the following multiple choice questions. First, we asked, ``How well, if at all, do each of the following describe this opinion statement? Aggressive, Informative, Well articulated, Persuasive to you.'' The first three items put the rater in an objective mindset, and the last one about persuasiveness allows the reader to express their own opinion so that the next question would be approached with an open mind.  The next question specifically asked, ``How persuasive, if at all, do you think the opinion would be for someone undecided on this issue?'' The last question assesses how intellectual humility in text affects perceptions of the text’s author: ``How enjoyable, if at all, would you find a conversation with the MTurker who wrote this opinion?'' All questions had five-point likert scale answer choices.

In addition to these main outcomes, we asked survey reviewers to rate texts based on other latent features related to intellectual humility. We asked them, given a political argument, ``How well it acknowledges uncertainty from 0 (not at all uncertain) to 100 (very uncertain), how well it respects the other side from 0 (not at all respectful) to 100 (very respectful), and how well it acknowledges the author's limitations from 0 (no acknowledgment of limitations) to 100 (significant acknowledgment of limitations).''

As we will show in Section~\ref{sec:experimental_design}, our experimental design effectively handles latent confounding and ensures we can unbiasedly estimate average treatment effects of latent treatments. We reserve the full analysis of this data for Section~\ref{sec:detailed_analysis} and use these outcomes to benchmark text-as-treatment estimators in Section~\ref{sec:sims}.

%% file: simulations.tex
Because the data we collected and analyze in Section~\ref{sec:detailed_analysis} operationalizes our experimental design, this data provides a high quality test bed for studying text as treatment estimators, including language model estimators. We leverage this data to generate a synthetic confounded dataset where we control the strength of confounding. Importantly, in all experiments, the text we used was generated by real participants and so this mimics a ``real world dataset'' experiment rather than synthetically generated texts, which would not necessarily have any human-interpretability. 

As discussed in Section~\ref{sec:ih_outcomes}, we asked survey respondents to rate their perceptions of our texts on a variety of outcomes \textit{and} other latent text features, such as the respectfulness of the text. We use this known latent feature of respectfulness to control the probability of selecting a given text to be part of a filtered dataset, therefore simulating selection bias. We examine two scenarios: in the first, referred to as the ``Baseline Confounding'' case, we use the observed outcomes without directly modifying them for respectfulness; in the second, the ``Amplified Confounding'' case, we systematically adjust outcomes to strengthen respectfulness as a latent confounder. Specifically, in the Amplified Confounding case, we adjusted the original, likert-scale outcomes--increasing them for respectful texts and decreasing them for disrespectful ones--to ensure respectfulness was predictive and acted as a confounder. To establish a non-zero treatment effect, we additionally adjusted continuous outcomes, increasing them for treated texts and decreasing them for untreated texts. Given that our data utilizes a five-point Likert scale but that \citet{pryzant2020causal}'s LLM algorithm under study only processes binary outcomes, we dichotomized our outcomes, reserving the full data analysis for Section~\ref{sec:detailed_analysis}. We generated and analyzed 100 replicas of this semi-synthetic dataset to assess estimator performance.

\subsection{Estimators}\label{sec:estimators}
We evaluated two naive estimators that do not account for latent textual features: Difference in means, which computes the simple difference in outcomes between treated and control texts, and Topic adjustment, which adjusts outcomes based solely on the text's topic (e.g., immigration or gun control). Additionally, we considered five advanced estimators that incorporate textual features. We employed a bag of words approach to structure the text for outcome regression (BoW OR), inverse propensity weighting (BoW IPW), and augmented inverse propensity weighting (BoW AIPW), using cross-fitted random forests to estimate all nuisance estimators. We also tested two language model estimators—TextCause, which treats the neural network as an outcome regressor \citep{pryzant2020causal}, and Treatment Ignorant (TI), which estimates conditional outcomes with a language model and trains propensity score model using the fitted language model's representations, integrating these via augmented inverse propensity weighting (AIPW) \citep{gui2022causal}. Due to extreme propensity scores rendering AIPW estimates non-computable, we applied winsorizing (TI Winsorized) and trimming (TI Trimmed) to ensure all scores ranged between 0.1 and 0.9 \citep{crump2009dealing}.

\subsection{Results}
\begin{figure}
    \centering
    \includegraphics[width=0.8\linewidth]{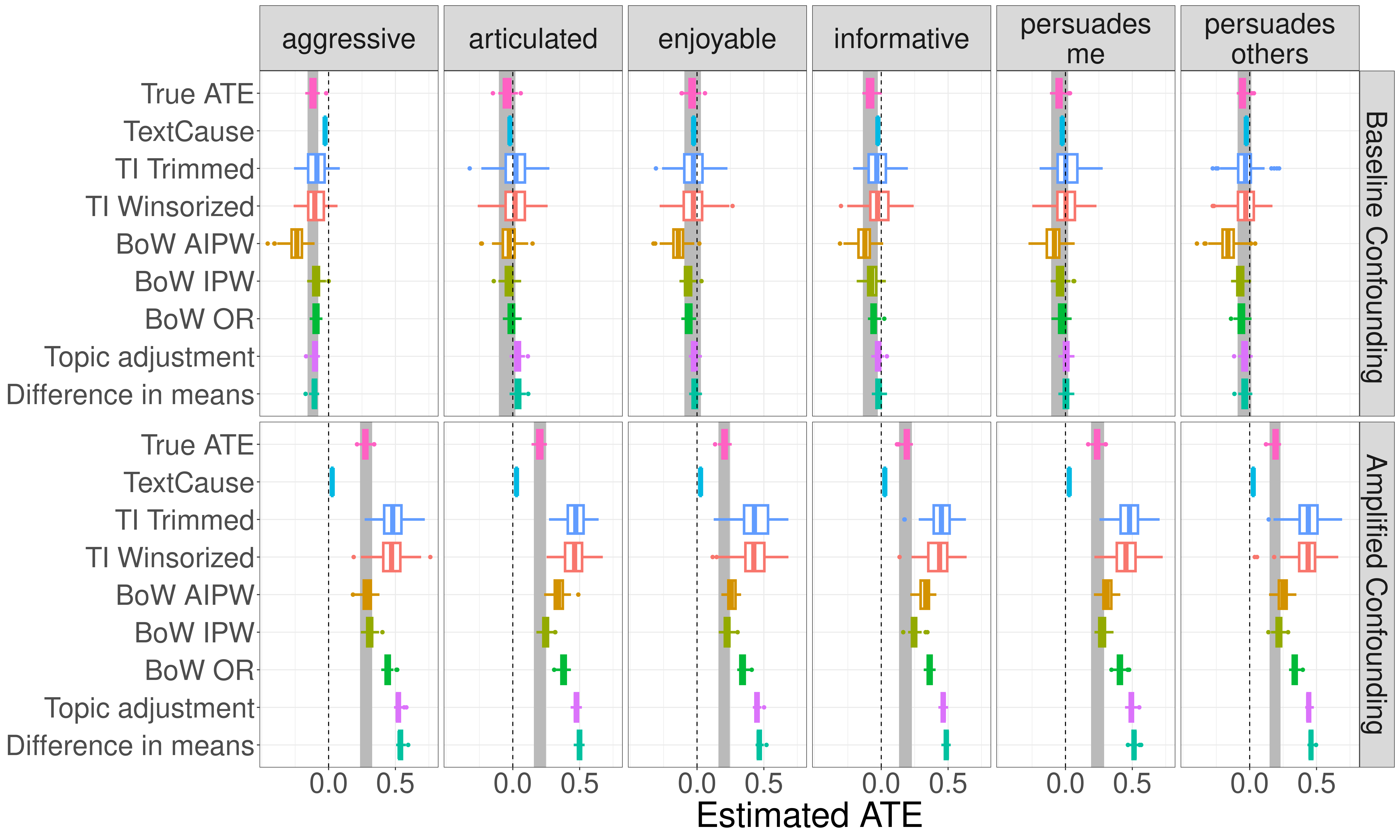}
    \caption{The top (bottom) panel shows estimates computed on the filtered datasets with the original (confounded) outcomes. We estimated the true average treatment effect (ATE) as described in Proposition 1. The 2.5 to 97.5 percentiles of these estimates across 100 replicas are shown as gray bands in each subplot. Each subplot features boxplots of the ATE estimated by each method for the 100 data replicas. The facet columns indicate the outcome label investigated in each experiment.}
    \label{fig:sim_results}
\end{figure}

Figure~\ref{fig:sim_results} presents our simulation results. The top (bottom) panel shows estimates computed on the filtered datasets with the original (confounded) outcomes. The gray bands in each subplot represent the 2.5 to 97.5 percentiles of estimates from the estimator in Proposition~\ref{prop:estimator}, serving as our ground truth ATE. Each subplot features boxplots of the ATE estimated by each method for the 100 data replicas. The facet columns indicate the outcome label investigated in each experiment.

In the Baseline Confounding case, \textit{almost all} of the baselines correctly capture the null effect of IH on the binarized outcomes. Because the naive difference in means estimate is also aligned with the ground truth, we conclude that there is little latent confounding in this text, suggesting that all estimators can achieve the right answer when there is both no effect and no latent confounding. Importantly, even though there is no latent confounding, the TextCause estimator still fails to correctly estimate the treatment effect, as seen in the \texttt{aggressive} panel of the top row.

The story further changes when we induce confounding (the bottom row). First, notice that all of the difference in means and topic adjustment estimates across all of the outcome labels are mis-estimated, validating the need for text adjustment to overcome omitted variable bias. Adjusting for non-latent text features (i.e., topic) is an insufficient representation of the rich latent confounding in this setting.

The much simpler bag of words representation with non-parametric estimators can estimate the true ATE far more accurately than the competing language model estimators. The IPW estimates are consistently inside the true ATE bands (gray), suggesting that \textbf{simple text representations can effectively represent confounding information in this case study}. However, all of the TI and TextCause algorithms fail to properly estimate the true treatment effect, which we will demonstrate stems from induced positivity violations.

TextCause consistently estimates null effects, which could be explained by two possibilities. First, the learned representations may have failed to capture any meaningful confounding information. The other possibility is that TextCause learns an embedding space that encodes the treatment itself, forcing the treated and control units to distinct parts of the latent space. In doing so, it is unable to make apples-to-apples comparisons between treated and control observations and therefore cannot properly impute a distinct counterfactual for each observation. Instead, it learns the same potential outcome for the treated and control observations, thereby forcing a null effect. 

If the estimator truly learned no meaningful confounding information, then the estimated effects would likely be more varied, with estimates also resembling those from the naive difference in means estimator. Because it has not, we can confidently conclude that the consistent null effect is not solely attributable to an inability to learn confounding information. \textbf{The TextCause algorithm encodes the treatment itself in the learned representations, thereby forcing a positivity violation.}

\input{tables/extreme_ps_estimates}

Because the TI estimator is an AIPW estimator, we can directly access the propensity scores it uses for estimation and diagnose overlap directly. Table~\ref{tab:extreme_ps_estimates} shows the distribution of estimated propensity scores for each outcome in one dataset of the complete confounding simulation. Across all of the outcomes, estimated propensity scores extended beyond 0.1 and 0.9, despite the fact that we \textit{designed} the simulations and data generation to ensure that positivity was satisfied. In contrast, propensities scores estimated with a bag of words text representation were between 0.1 an 0.9 for $83\%$ of observations for all outcomes (as seen in Table~\ref{tab:extreme_ps_estimates_bow}). Other runs have similar distributions of propensity scores, highlighting how these language models can induce positivity violations.

All 100 replicates for each of the six outcomes in the complete confounding simulation have at least one observation with an estimated propensity score of 0 or 1, making the AIPW estimate non-computable. Instead, we employed both propensity score trimming and winsorizing so that units have estimated propensity scores between 0.1 and 0.9 \citep{crump2009dealing}. As seen in Figure~\ref{fig:sim_results}, the TI Trimmed and TI Winsorized estimates completely fail to estimate the true treatment effect. Because, the TI Trimmed and TI Winsorized estimates are strikingly similar, it is unlikely that the trimmed population \citep{crump2006moving} is too different than the original population. Instead, notice that the TI Trimmed and Winsorized estimates resemble the Difference in means estimates, suggesting that the learned representations fail to capture any meaningful confounding information. The latent space may have instead encoded effect modifiers that are highly predictive of the outcome but not the treatment, and the treatment itself.

\paragraph*{Discussion} Our simulation studies highlight key issues in current approaches for estimating causal effects with text as treatment. First, bigger is not always better. Even though language models are more powerful than simple text representations, they may learn the treatment itself, inducing positivity violations. 

Second, language models are not guaranteed to learn any meaningful confounding information either; they may instead learn highly predictive effect modifiers. It is important to note that our simulation environment is deliberately constructed to be favorable to LLM-based, covariate adjustment estimators. This setup is so simple that even a basic bag-of-words estimator can accurately recover the true treatment effect. Therefore, the failure of LLM-based estimators in this environment is not due to adversarial construction, but rather highlights a fundamental limitation in their ability to preserve causal identification even under extremely forgiving conditions.

Lastly, simpler representations can, but not always will, capture confounding information properly. While the BoW IPW almost always correctly estimates the true treatment effect, BoW OR fails to estimate the ground truth accurately; because model selection is so difficult and challenging in causal inference, knowing which estimator produces the most reliable results requires strong assumptions \citep{parikh2022validating, rudolph2023all, van2024can}. Rather than relying on model-based adjustments to study the causal effects of linguistic properties, we instead introduce a novel experimental design that is (1) simple to understand, (2) guaranteed to produce accurate results, and (3) requires very mild assumptions.

%% file: tables/extreme_ps_estimates.tex

\begin{table}[t] \centering 
\begin{singlespace}
  \caption{The proportion of observations for a single run of each confounded outcome with extreme propensity scores estimated using the TI Estimator.} 
  \label{tab:extreme_ps_estimates} 
\begin{tabular}{@{\extracolsep{5pt}} lccc} 
\\[-1.8ex]\hline 
\hline \\[-1.8ex] 

Outcome & Est Prop $\leq 0.1$ & $0.1 <$  Est Prop $<  0.9$ & $0.9 \leq$ Est Prop \\ 
\hline \\[-1.8ex] 
\texttt{aggressive} & 0.08 & 0.67 & 0.25 \\ 
\texttt{articulated} & 0.1 & 0.79 & 0.11 \\ 
\texttt{enjoyable} & 0.17 & 0.71 & 0.12 \\ 
\texttt{informative} & 0.1 & 0.77 & 0.14 \\ 
\texttt{persuades me} & 0.08 & 0.85 & 0.07 \\ 
\texttt{persuades others} & 0.07 & 0.83 & 0.1 \\ 

\hline \\[-1.8ex] 
\end{tabular} 
\end{singlespace}
\end{table}

%% file: experimental_design.tex
This section is organized as follows. First, we specify the causal estimand we and the other methods in Section~\ref{sec:sims} seek to estimate. Then, we describe existing methodology to estimate them, highlighting places where strong assumptions need to be made. Finally, we develop a procedure to assess the relevant causal effects with careful design of the text and randomization. This procedure allows us to control the level of confounding while still having access to unbiased ground truth with real data. We can then validate the model-based causal estimates by comparing them to the experimental results. 

\subsection{Causal Estimands for Text}

\begin{figure}[ht!]
    \centering
    
    \begin{tikzpicture}
    
    \node[obs,xshift = -1in] (D) {$D$};%
    \node[obs] (W) {$W$};%
    \node[obs, xshift = 2in] (Y) {$Y$};%
    \node[latent,above= of W,xshift = 1in] (Z) {$Z$};%
    \node[latent,below= of W,xshift = 1in] (T) {$T$};%
    
    \edge{D}{W}
    \edge{W}{Z}
    \edge{Z}{Y}
    \edge{W}{T}
    \edge{T}{Y}

    \end{tikzpicture}
    \caption[Causal diagram for causal effects of text]{ Causal diagram of document labels for the latent treatment $D$, words $W$, the latent treatment itself $T$, other latent content $Z$, and outcome $Y$. Throughout this section, we assume that $D$ and $T$ are binary. Shaded nodes are observed, non-shaded nodes are unobserved but (potentially) measurable from $W$. There are three relevant potential outcomes that are equivalent under the standard SUTVA assumption \citep{rubin1980randomization}. First, $Y_i(D)$ is unit $i$'s outcome depending on assigned document label $D$. Second, $Y_i(W) = Y_i(W(D))$, the outcome depending on the words in the document read by unit $i$, which are determined by $D$. Third, $Y_i(T,Z) = Y_i(T(W),Z(W))$, the outcome dependent on the latent features in the words, which are of course determined by the words themselves.}
    \label{fig:text_dag}
\end{figure}
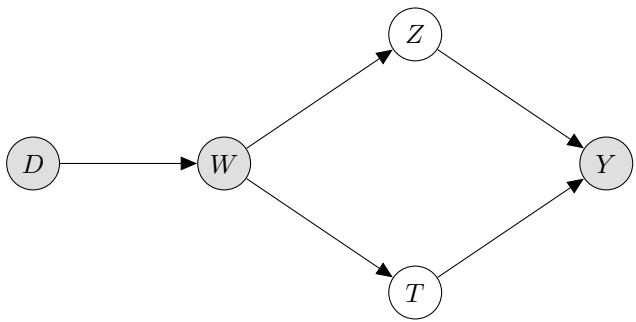

Before detailing the relevant causal estimands, we need to clarify some assumptions about the relevant quantities. Figure~\ref{fig:text_dag} shows a causal directed acyclic graph (DAG) that represents the causal relationships between the following: a known document label $D$, which indicates whether some latent feature of interest (treatment $T$) is present in a document, the words in that document $W$; the outcome-relevant, latent treatment $T$; outcome-relevant, latent features $Z$; and the outcome of interest $Y$. The implication of this construction is that the physical words on a page or pixels on a screen do not determine the outcome. Rather, the language communicates some information ($T$ and $Z$) to the reader, and it is those latent concepts that determine the outcome. In potential outcome notation, this relationship can be expressed as $Y(W=w) = Y(T(W=w),Z(W=w))$. The nested potential outcome notation clarifies that the potential outcome for a reader when assigned $W=w$ is identical to the potential outcome when that reader is assigned a $W=w'$ that has the same $T$ and $Z$. 

A few additional assumption about the Figure~\ref{fig:text_dag} DAG are commonly made. We also make those assumptions and enumerate them here: 
\begin{enumerate}
    \item \underline{No misclassification}. $D=T$, the known document labels correspond to the actual latent treatment, i.e. whatever classification method was used is accurate.\footnote{In \citet{gui2022causal} and \citet{pryzant2020causal}, a slightly different DAG is used where $D$ represents the author's intentions, $T$ the reader's perceptions, and $D=T$ is the assumption that these intentions and perceptions align.} While potentially a strong assumption, this one is easily validated by examining observed document content $W$ and document labels $D$. For potential outcomes, this means that $Y(D,T,Z) = Y(T,Z) = Y(D,Z)$. The potential outcome is fully specified for a given $Z$ with the addition of either $T$ or $D$. 
    \item \underline{Consistency of interpretation}. $T$ and $Z$ are deterministic functions of $W$. In the potential outcome notation used above, this means that $Pr(T(W) = t|W) \in \{0,1\}$. This assumption is often reasonable, as a fixed set of words has a fixed meaning in the population of interest. 
\end{enumerate} 
The implication of (1) and (2) is that $D$ also must be a deterministic function of $W$, but we usually do not think of it that way because in our representation $D$ is causally ``upstream'' of $W$. In a hypothetical experiment, researchers might randomly assign participant $i$ to either $D_i=1$ (reading text with the treatment of interest) or to $D_i=0$ (reading text without the treatment of interest), which then determines a second random assignment of drawing $W_i$ from either a pool of documents with the feature ($D_i=1$) or from a disjoint pool of documents without the feature ($D_i=0$). An implication of $(2)$ alone is that no confounding of the $Z\to Y$ or $T\to Y$ relationship is possible. A confounder would need to be an additional, non-$W$ variable that affects $Z$ or $T$. Such variables cannot exist because $W$ exactly determines the latent concepts represented in the text. If one relaxes $(2)$ to allow different people to have different interpretation of the document content, this kind of confounding would be possible. 

Now, we return to characterizing the causal effects of interest. There are two kinds of effects researchers are interested in when it comes to measuring the causal effect of text. The first and most common is the effect of exposure to \emph{documents with some latent feature $T$}. In the DAG in Figure~\ref{fig:text_dag}, this is represented in the effect on $Y$ from intervening on $D$. The second effect of interest is the effect of exposure to \emph{the latent feature itself.} This is the effect of intervening on $T$. We start by describing the first effect. 

Let $\tau_d = E[Y_i(D=1) - Y_i(D=0)] = E[Y_i(T=1,Z=z) - Y_i(T=0,Z=z')]$, where the expectation is taken over $i$, the units in the population, and $Z$ is not held fixed. This is the effect of exposure to documents with some latent feature. Note that this effect comes both from the fact that $T$ varies with $D$ and that (potentially) $Z$ also varies with $D$, going from $z$ to $z'$. Estimating $\tau_d$ with a randomized experiment is straightforward. Collect some documents, measure the treatment of interest, and randomly assign participants to read either documents with or without the treatment feature. Estimating the second effect, even with randomization, is much more challenging. 

Let $\tau_t = E[Y_i(T=1) - Y_i(T=0)] = E[Y_i(T=1,Z=z) - Y_i(T=0,Z=z)]$. This is the effect of varying only the feature of interest, holding $Z$ fixed. This effect is more difficult to conceptualize. In theory, we would want to intervene on the above DAG to set $T$ to some level without changing anything causally ``upstream'' of $T$, i.e. $W$. However, this is clearly impossible: changing a message from polite to impolite, for example, requires changing the actual words in the message, which can induce changes in $Z$. An easier way to understand this effect is to consider the decomposition of $\tau_d$ as shown below (expectations over $i$ are omitted for clarity): 
\begin{align}
    \tau_d &= E[Y(D=1) - Y(D=0)] \\
    &= E_Z[Y(D=1,T=1,Z=Z(D=1)) - Y(D=0,T=0,Z=Z(D=0))] \\
    &= E_Z[Y(D=1,T=1,Z=Z(D=1)) - Y(D=1,T=1,Z=Z(D=0)) \label{eq:add_zero_1}\\
    & \ \ \ + Y(D=1,T=1,Z=Z(D=0)) - Y(D=0,T=0,Z=Z(D=0))] \label{eq:add_zero_2}\\
    &= E_Z[Y(D=1,T=1,Z=Z(D=1)) - Y(D=1,T=1,Z=Z(D=0))] + \tau_t. \label{eq:taud_decomp_final} 
\end{align}

The second line takes an expectation over the non-treatment features $Z$. The notation $Z(D)$ indicates that the non-treatment features can also vary with the treatment feature. An expectation is not needed over $T$ because $T=D$ by assumption. Note that we include the $D$ term in the potential outcome notation only for clarity of exposition because the effect is entirely captured via $Z$ and $T$; it clarifies that we are never assessing a potential outcome where $D\neq T$. Equations \ref{eq:add_zero_1} and \ref{eq:add_zero_2} add and subtract the same quantity, then Equation \ref{eq:taud_decomp_final} recognizes that \ref{eq:add_zero_2} is the causal effect of changing $T$ without changing $Z$. In essence, we have decomposed $\tau_d$ into the effect on $Y$ that goes through $Z$ and the effect that goes through $T$. 

The two effects provide important insights that can guide different decision making tasks. A platform that recommends text to users, such as a social media company organizing a news feed, should be more interested in $\tau_d$. It does not matter to them why posts with some feature are more engaging, only that the engagement is a function of an observable label. On the other hand, authors who are trying to optimize a specific message are often much more interested in $\tau_t$. They have full control over the actual text and can choose each message feature to produce a result. 

\section{Estimating $\tau_t$}\label{sec:estimating_taut}

A naive examination of the Figure~\ref{fig:text_dag} DAG could lead one to think that estimating $\tau_t$ is straightforward by simply conditioning on the observed variable $W$. This conditioning would require comparing the outcomes for two units with the same $W$ but different $T$. However, this comparison is clearly impossible due to the underlying concepts the DAG represents. It is impossible for two documents to have the same words and differ on the treatment feature. This is a violation of the overlap condition: $Pr(T=1|W=w) \in \{0,1\}$. 

Several recent papers attempt to overcome this issue by developing methods that, under certain additional assumptions, can identify $\tau_t$ \citep{pryzant2020causal, gui2022causal, fong2023causal}. These methods fall into one of two categories: (1) covariate adjustment that attempts to learn $Z$ through some language model or (2) careful curation of texts such that $Z$ is independent of $T$. As we demonstrate in Section~\ref{sec:sims}, existing covariate adjustment methods can induce overlap violations. In this section, we further expand on text-curation approaches and present our experimental design.

\subsection{Curation of Texts}

This approach is developed in \citet{fong2023causal} and expanded upon below. The key observation is if $Z$ is independent of $T$ (and $D$ by trivial extension), then the expectation in Equation~\ref{eq:taud_decomp_final} is 0. The potential outcomes $Z_i(D_i=1)$ and $Z_i(D_i=0)$ are equal in distribution over $i$. This can hold when the researcher has carefully designed their own texts or curated naturally occurring text such that for every text with latent features $(T=1,Z=z)$ there exists another text with features $(T=0,Z=z)$ and the assignment of texts to participants is equally likely to select either text. 

The main limitations are quite clear: creating or finding such texts can be challenging because $Z$ is typically not known. If the researcher can generate texts themselves, they can produce documents that vary on $T$ but hold $Z$ constant.  For example, \citet{fong2023causal} design messages about protests in Hong Kong by randomly choosing from a large population of potential message features from an existing corpus of texts. We leverage a similar setup that uses study participants, rather than an existing observational corpus, to produce natural texts themselves that vary on a treatment feature, then edit the messages to change treatment while holding all else fixed. This gives us much wider and more natural variation in the expression of the latent treatment and other confounding features. 

The four-step procedure is outlined as follows: first, choose some linguistic feature of interest (treatment $T$) and at least two topics (potential confounding $Z$s). Second, ask respondents to generate texts with ($T=1$) or without ($T=0$) the linguistic feature and about one of the topics. We refer to this group as the writers. Third, recruit a different set of respondents and have them edit the texts generated in the previous step to change $T$ but leave everything else the same. Each original text should have multiple edits. We refer to this group as the editors. Fourth, recruit a final set of respondents to read each text and evaluate it for some outcome metric that $T$ and topic could potentially causally influence. We refer to this group as the evaluators. The following details each step and discusses potential considerations. 

\begin{enumerate}
    \item Step One: Choosing Features, Topics, and Outcomes. Generally, applied researchers will already have a specific feature and topics in mind, so this step is often straightforward. The important considerations are that the pool of study participants should already understand or be able to learn how to write text with and without the feature that is topic relevant.

    \item Step Two: Generating Texts. The goal of this step is to generate many messages that vary on both treatment and the topic. One practical consideration is that we may want a single respondent to generate several texts. Many language models do require large amounts of data, and treatment effects could be quite small. If the treatment or topic are complex issues that respondents need to be trained on, budgetary constraints can become relevant too. After spending time learning about the concepts and how to generate texts, the actual generation of additional texts is quite quick. Because outcomes in Step Four are generated by separate respondents looking at only the text, there is little concern about inducing dependence across responses. Even if it were a concern, the researcher at least knows who authored which texts and what texts were shown, which allows for relevant downstream adjustment. 

    \item Step Three: Editing Texts. This step likely requires the same training as the previous step on how to write text with and without the treatment feature. Additionally, it is important that not just the topic is preserved with the edit, but also everything besides treatment itself. If, for example, respondents are asked to write a message about gun control (the topic $Z$), and the original text discusses background checks, the edited text should still also discuss background checks. Importantly, a text can only be included in the analysis if it has a valid edit. By collecting multiple edits for the same original text, researchers can ensure that respondents failing to edit properly will not waste the work done writing the text. Multiple edits can also allow the researcher to induce confounding, as discussed in Section~\ref{sec:analysis_procedure}. 

    \item Step Four: Outcome Generation. Again, applied researchers may already have a clear outcome of interest. The only statistical consideration here is that, as noted previously, for comparisons between methods it is helpful  if the treatment has an effect on the outcome. It is also very important that the topic has an effect on the outcome because the topic will be used as a confounder, and a confounder must affect the outcome. 
\end{enumerate}

\subsection{Analysis Procedure}\label{sec:analysis_procedure}

The recommended estimator for $\tau_t$ is the difference in outcomes between the original text and its edits, averaged across all original texts. Below, let $Y_{ij}$ denote the outcome for the $j$th version of original text $i$ ($j=1$ is the original text), $J_i$ is the number of versions of original text $i$, and let $T_{ij}$ be the treatment indicator for text $ij$. Note that for $j>1$, $T_{ij} = 1-T_{i1}$.  Thus the estimator is given by:  
\begin{align}
    \widehat{\tau}_t :=\frac{1}{N} \sum_{i=1}^{N} \left(\frac{1}{\sum_{j=1}^{J_i} T_{ij}} \sum_{j=1}^{J_i} Y_{ij} T_{ij}  - \frac{1}{\sum_{j=1}^{J_i} 1-T_{ij}} \sum_{j=1}^{J_i} Y_{ij} (1-T_{ij})  \right).\label{eq:taut_hat}
\end{align}
As long as the edits have preserved everything except the treatment, the effect of all possible confounders $Z$, not just the assigned topic, will be differenced out and $\widehat{\tau}_t$ will be an unbiased estimate of $\tau_t$. The estimator can be implemented via weighted least squares with weights proportional to the normalizing terms above, i.e. the weight for $Y_{ij}$ is $(\sum_{j'=1}^{J_i} \mathbf{1}\{T_{ij'} = T_{ij}\})^{-1}$, an indicator variable for treatment, and a fixed effect for each $i$. Inference can be performed either with permutation tests on the treatment indicator between the original and edited texts for each $i$ or with the asymptotic results for weighted least squares. This is formalized in the proposition below.  

\begin{proposition}  \label{prop:estimator}
Suppose that for each original text $W_{i1}$, the edited versions $W_{ij}$ ($j>1$) are such that $T(W_{i1}) = 1-T(W_{ij})$ and $Z(W_{i1}) = Z(W_{ij})$. Additionally, suppose that texts are randomly assigned to respondents, who generate the outcomes. Then, $\widehat{\tau}_t$ is an unbiased estimator of $\tau_t$ and can be identified with weighted least squares. Proof in Appendix~\ref{app:prop_proof}. 
\end{proposition}

We note that if each original text has exactly one edit, then in this sample, $Z$ and $T$ are independent. Knowing any $Z$, not just topic, provides no information about $T$ because the sample is exactly balanced by construction. In this case, Equation~\ref{eq:taut_hat} reverts to the simple difference in means between $Y_{ij}$ such that $T_{ij}=1$ and $T_{ij} = 0$ (the weights are uniform). This is exactly the estimator proposed in \citet{fong2023causal} when a researcher designs their text in a similar manner. 

Thus, differences in the number of edited documents, especially when that number is large only for original documents about a specific topic and treatment status, allow us to control the level of confounding in the full sample, when not accounting for $i$. Consider a hypothetical where all original texts received one edit except for original texts with $T_{i1} = 1$ and an assigned topic of gun control, which received four edits. In the full sample, without adjusting for the original text, knowing the topic is gun control is predictive of treatment. If the topic is gun control, then the probability that the document has $T_{ij} = 1$ is 35\%.\footnote{The population is 25\% originals with $T=0$, 25\% edits of those documents with $T=1$, 10\% originals with $T=1$ and 40\% edits of those documents with $T=0$.} For all other topics, the probability is 50\%. Under these conditions, the most relevant assumption for the covariate adjustment methods, overlap, is still met. For all non-treatment features $Z$, the probability of treatment is bounded away from 0 and 1. The remaining assumptions focus on whether the language model can recover appropriate proxies of $Z$, which may or may not hold. On the other hand, our estimator properly adjusts for confounding from any $Z$ regardless of population sizes, so it can always be used to provide a ground truth.

%% file: detailed_analysis.tex
In this section, we analyze the data collected using the experimental design in Section~\ref{sec:experimental_design}. We evaluate the causal effect of linguistic markers of intellectual humility on perceptions of a communicated political argument.

Our survey experiment operationalized the design discussed in Section~\ref{sec:experimental_design}. First, we trained writers to learn that intellectual humility involves (1) showing respect for others' beliefs and ideas, (2) considering counterpoints to one's own views, and (3) admitting limitations of uncertainties in one's own beliefs. Writers were assigned to two groups. Those in the first group were asked to share their opinions on an assigned topic --  gun control, climate change, or immigration -- while expressing intellectual humility. On the other hand, those assigned to the other group were given the same prompt but were asked to share opinions while \textit{not} being intellectually humble. This first stage yielded 176 IH and 167 non-IH texts.

\input{tables/text_edits}

We then trained another set of participants, editors, to learn about intellectual humility (see Section~\ref{sec:survey_items} for more details). We asked these participants to edit their assigned text's intellectual humility while preserving all other features. Each original text had at least one edited text, and both original and edited texts had multiple evaluations. Table~\ref{tab:text_edits} displays an edited version of the non-IH text displayed in Table~\ref{tab:text_examples}.Both the original and edited texts maintain colorful language, but differ in their expression of intellectual humility (IH), demonstrating the strength of our experimental design and editorial process.

\begin{figure}
    \centering
    \includegraphics[width=0.5\linewidth]{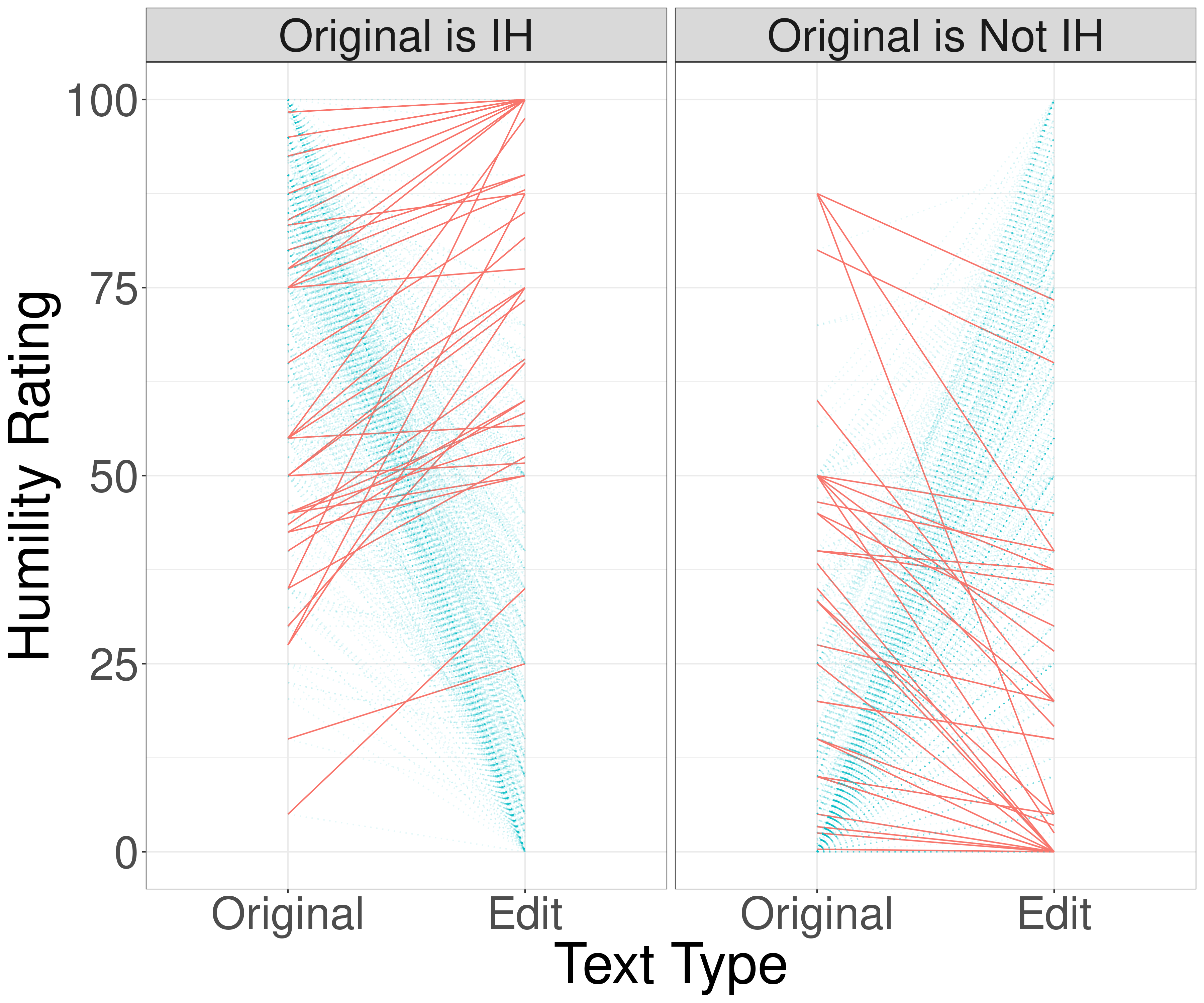}
    \caption{X axis displays whether a text was an original or an edit. The y-axis represents the IH rating of each text on a scale from 0-100. Each line links the IH rating for an (original, edit) pair. Texts that are originally IH should see a decreasing line since the edit should not be IH. In contrast, texts that are originally not IH should see an increasing line since the edit should be IH. Lines that follow the expected slope are blue, dotted, and faded while those that are wrong are in red, solid, and bold.}
    \label{fig:ih_slopes}
\end{figure}

We then asked another set of participants, evaluators, to rate the texts on a 5-point likert scale for each of our outcomes of interest: how (1) aggressive, (2) articulated, (3) informative the text seemed, (4) how enjoyable it would be to converse with the text's author, (5) how persuasive the text was to the reader, and (6) how persuasive the text would be to others. We ultimately collected 6994 evaluations of 1830 unique texts from 1400 evaluators.

We also leveraged these evaluators to audit our  editors' performance. After an evaluator rated her texts on each of the six outcomes, she also learned about intellectual humility. She then rated the intellectual humility of another set of texts on a scale from 0-100. Figure~\ref{fig:ih_slopes} displays each text's IH rating, averaged across evaluators, with the lines connecting (original, edit) pairs. The left panel displays the ratings for texts that were originally IH while the right panel displays ratings for texts that were originally not IH. Solid red lines in Figure~\ref{fig:ih_slopes} denote edited texts in which humility moved against the instructed direction; these texts were subsequently removed from our corpus ($5.95\%$). We subsequently removed $4.74\%$ of original texts that had no remaining edited pairs after removing bad edits. Our final sample consists of 1682 texts and, for each outcome, 6195 evaluations of the texts. Table~\ref{tab:exp_ss} displays a break down of the number of evaluations by topic, IH, and original/edit status.

\input{tables/real_data_sample_size}
We estimated the average treatment effect using our weighted least squares estimator introduced in Proposition~\ref{prop:estimator}. To account for dependencies between a single evaluator rating multiple texts, we also included evaluator-level random effects in the weighted least squares analysis. 

\begin{figure}[t]
    \centering
    \includegraphics[width=0.5\linewidth]{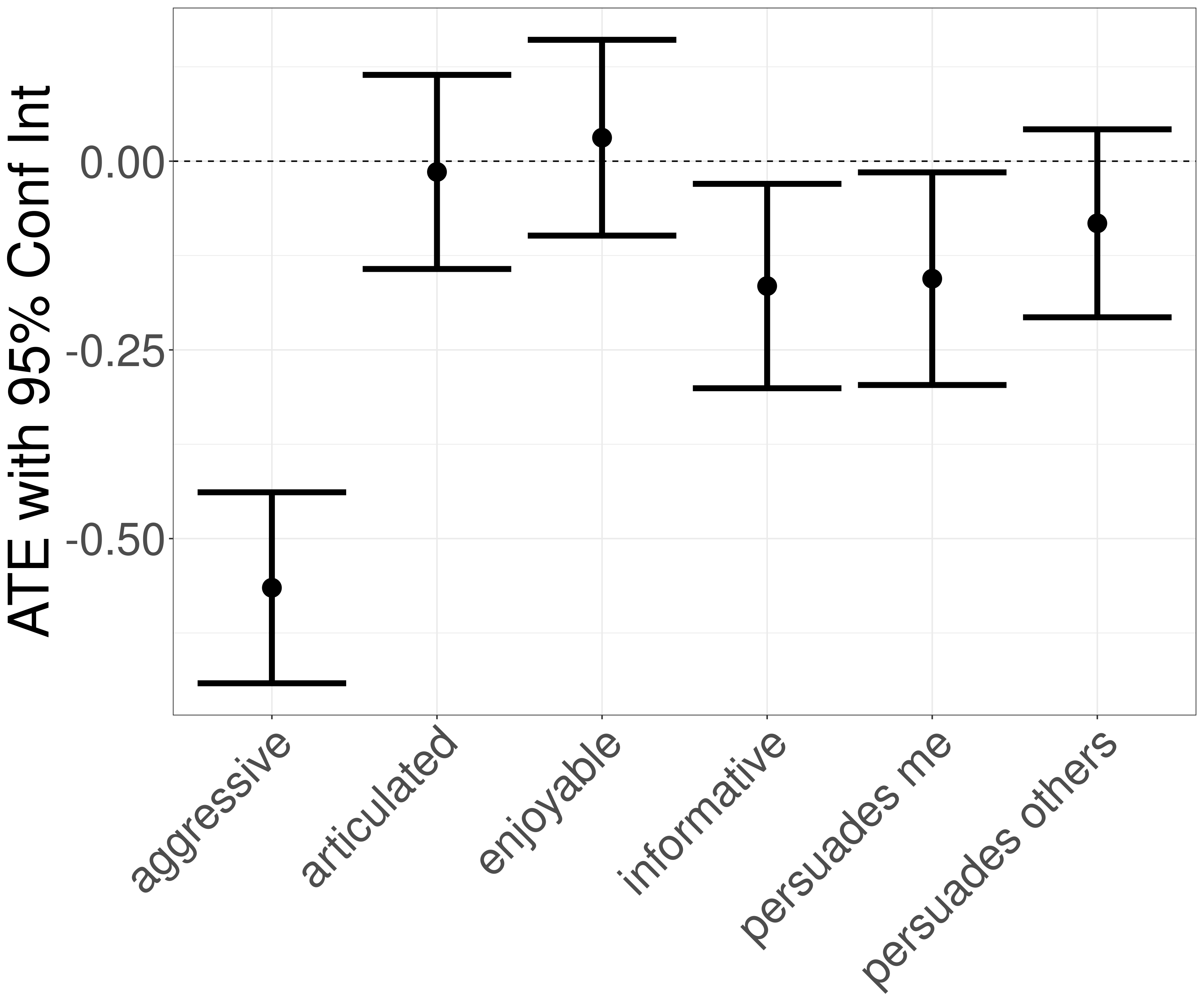}
    \caption{95\% confidence intervals (y-axis) showing the average treatment effect of intellectual humility on each outcome (x-axis). The dotted line represents no effect at 0.}
    \label{fig:exp_results}
\end{figure}

As seen in Figure~\ref{fig:exp_results} and Table~\ref{tab:real_data_regs}, we find that the use of intellectually humble language in a political communication decreases perceived aggressiveness, consistent with previous research that finds intellectual humility promotes positive political exchange (e.g., \citet{knochelmann2024effects}).\footnote{Please note that LLM benchmarks would not work for our data because they can only take as input one non-text confounder (e.g., \citet{gui2022causal}) or only binary outcomes (e.g., \citet{pryzant2020causal}).}  However, we find that intellectually humble language also causes a significant decline in the perceived informativeness and first-order persuasiveness of the communication. It also has no impact on how articulate the argument is rated, how enjoyable conversation might be, or on perceptions of how persuasive the text would be to others. Previous research had lauded the potential for intellectual humility to foster open and productive dialogue that would reduce polarization \citep{porter2018intellectual, montez2024effects, knochelmann2024effects}. 

Our results point to an apparent paradox for the use of intellectually humble language on the resulting communication outcomes. While intellectually humble language might soften perceptions of aggression—potentially fostering a more approachable and civil dialogue—it also potentially dilutes the effectiveness of the communication by reducing the perceived informativeness and persuasiveness. It seems that intellectually humble language might undermine the authority, conviction, and confidence of an argument, which could impact its persuasiveness \citep{perloff2017dynamics}. In many ways, this pattern is reminiscent of the empirical literature on negative advertising, which consistently finds that people very much dislike a negative tone, but that it is nonetheless effective (e.g., \citet{dowling2016effects, zhang2024negative}).

\input{tables/real_data_regs}

%% file: tables/text_edits.tex
\begin{table}[t]
\centering
\caption{The results from our experimental design: the originally \arrogant{intellectually arrogant} argument from Table~\ref{tab:text_examples} is displayed in the first column. The \textbf{edited version} of the same text that preserves everything -- including the \colorful{colorfulness} of the language -- \textit{except for} \humble{intellectual humility} is displayed in the right column. Our experimental design effectively handles latent confounding without needing a black box language model.} \label{tab:text_edits}
\begin{tabularx}{\textwidth}{X|X}
\hline 
\textbf{Original} \arrogant{intellectually arrogant} argument \color{black} &  \textbf{Edit} that is \humble{intellectually humble} \color{black}  \\
\hline
Gun control \arrogant{needs to be} passed immediately. \arrogant{I know} that guns are \colorful{weapons of murder} that need to be controlled. Guns \arrogant{do not} protect people. & 
\humble{I think} that gun control should be passed soon. I believe that guns \humble{can be} \colorful{weapons of murder} and it would be better if they were controlled. \humble{I'm no expert}, but \humble{I don't think} that guns protect people.\\
\hline
\end{tabularx}
\end{table}

%% file: tables/real_data_sample_size.tex
\begin{table}[!htbp] \centering 
  \caption{Number of evaluated texts in each category before and after filtering. Topic describes the political issue respondents were told to write about. IH refers to whether intellectual humility was expressed in the final version of the text. Original indicates whether the text was an original creation or edited.} 
  \label{tab:exp_ss} 
\begin{tabular}{@{\extracolsep{5pt}} lcccc} 
\\[-1.8ex]\hline 
\hline \\[-1.8ex] 
Topic & IH & Original & Raw \# of evaluations & Final \# of evaluations \\ 
\hline \\[-1.8ex] 
Climate Change & No & No & 1332 & 1264 \\ 
Climate Change & No & Yes & 344 & 342 \\ 
Climate Change & Yes & No & 1259 & 1176 \\ 
Climate Change & Yes & Yes & 503 & 503 \\ 
 & & & & \\
Gun Control & No & No & 688 & 648 \\ 
Gun Control & No & Yes & 154 & 151 \\ 
Gun Control & Yes & No & 609 & 576 \\ 
Gun Control & Yes & Yes & 248 & 174 \\ 
& & & & \\
Immigration & No & No & 571 & 530 \\ 
Immigration & No & Yes & 154 & 151 \\ 
Immigration & Yes & No & 545 & 516 \\ 
Immigration & Yes & Yes & 164 & 164 \\ 
\hline \\[-1.8ex] 
\end{tabular} 
\end{table} 

%% file: tables/real_data_regs.tex
\begin{table}[!htbp] 
\centering 
  \caption{Regression Results for Real Data Analysis} 
  \label{tab:real_data_regs} 
\footnotesize
\begin{tabular}{@{\extracolsep{1pt}}lD{.}{.}{-2} D{.}{.}{-2} D{.}{.}{-2} D{.}{.}{-2} D{.}{.}{-2} D{.}{.}{-2} } 
\\[-1.8ex]\hline 
\hline \\[-1.8ex] 
 & \multicolumn{6}{c}{\textit{Dependent variable:}} \\ 
\cline{2-7} 
\\[-1.8ex] & \multicolumn{1}{c}{aggressive} & \multicolumn{1}{c}{articulated} & \multicolumn{1}{c}{enjoyable} & \multicolumn{1}{c}{informative} & \multicolumn{1}{c}{persuades (self)} & \multicolumn{1}{c}{persuades (other)} \\ 
\\[-1.8ex] & \multicolumn{1}{c}{(1)} & \multicolumn{1}{c}{(2)} & \multicolumn{1}{c}{(3)} & \multicolumn{1}{c}{(4)} & \multicolumn{1}{c}{(5)} & \multicolumn{1}{c}{(6)}\\ 
\hline \\[-1.8ex] 
 Const & 2.11^{***} & 2.46^{***} & 1.84^{***} & 2.35^{***} & 2.09^{***} & 2.31^{***} \\ 
  & (0.30) & (0.29) & (0.26) & (0.27) & (0.29) & (0.26) \\ 
  & & & & & & \\ 
 IH & -0.57^{***} & -0.01 & 0.03 & -0.17^{**} & -0.16^{**} & -0.08 \\ 
  & (0.06) & (0.07) & (0.07) & (0.07) & (0.07) & (0.06) \\ 
  & & & & & & \\ 
\hline \\[-1.8ex] 
\#Obs & \multicolumn{1}{c}{6,195} & \multicolumn{1}{c}{6,195} & \multicolumn{1}{c}{6,195} & \multicolumn{1}{c}{6,195} & \multicolumn{1}{c}{6,195} & \multicolumn{1}{c}{6,195} \\ 
LogLik &  \multicolumn{1}{c}{-9,306.73} & \multicolumn{1}{c}{-9,277.19} & \multicolumn{1}{c}{-8,744.47} & \multicolumn{1}{c}{-8,969.41} & \multicolumn{1}{c}{-9,271.09} & \multicolumn{1}{c}{-8,675.30} \\ 
AIC &  \multicolumn{1}{c}{19,547.46} & \multicolumn{1}{c}{19,488.37} & \multicolumn{1}{c}{18,422.95} & \multicolumn{1}{c}{18,872.81} & \multicolumn{1}{c}{19,476.18} & \multicolumn{1}{c}{18,284.60} \\ 
BIC &  \multicolumn{1}{c}{22,691.07} & \multicolumn{1}{c}{22,631.98} & \multicolumn{1}{c}{21,566.55} & \multicolumn{1}{c}{22,016.42} & \multicolumn{1}{c}{22,619.79} & \multicolumn{1}{c}{21,428.21} \\ 
\hline 
\hline \\[-1.8ex] 
  & \multicolumn{6}{r}{$^{*}$p$<$0.1; $^{**}$p$<$0.05; $^{***}$p$<$0.01} \\ 
\end{tabular} 
\end{table}

%% file: conclusion.tex
We introduce a novel experimental design for causal inference with latent treatments that addresses both latent confounding and the positivity assumption. Our design permits unbiased estimation of the average treatment effect, thereby providing a valuable testbed for evaluating existing text-as-treatment estimators. To benchmark these estimators, we implement a survey experiment operationalizing our design to study the effect of expressing intellectual humility on perceptions of political arguments. Empirically, we find that bigger is not always better: language model-based estimators may induce violations of the positivity assumption in the embedding space, leading to biased estimates, while simpler text representations can still accurately estimate treatment effects.

From a substantive perspective, our results reveal a tension in the communicative effects of intellectually humble language: while such language may attenuate perceived aggression, it may also reduce perceived authority, conviction, or confidence--thereby undermining communicative efficacy. These findings suggest that intellectual humility, though well-intentioned, may not serve as a straightforward intervention for reducing dogmatism or polarization in political discourse, particularly in online environments.

\paragraph*{Limitations and Future Directions}
Our design is an important step towards enabling causal inference in the context of linguistic treatments. While it facilitates estimation of causal effects in the domain of isolated statements, many relevant research questions pertain to conversations, where linguistic features interact dynamically. For example, although we find that intellectually humble statements are less persuasive than their non-humble counterparts, it remains unclear how humble language influences outcomes in interactive, conversational settings. Future work should develop designs capable of capturing the causal effects of communication in conversational contexts. 

In addition, experimental studies are not always feasible due to ethical, logistical, or practical constraints. Therefore, continued methodological innovation is needed to support retrospective causal inference with latent treatments for observational data analysis. As our results underscore, such methods must address both confounding and overlap more carefully. Our data and procedure offer a benchmark for the validation of new estimators in this domain.

%% file: appendix.tex
\subsection{Baseline Implementations}
\paragraph{TextCause \citep{pryzant2020causal}}
TextCause employs a pre-trained language model, specifically DistilBERT, to generate text representations. These representations are then used to predict the observed treated outcomes and the observed control outcomes, essentially capturing the confounding elements within the text that are associated with the linguistic properties being studied. This is akin to an outcome regression estimator. 

The representations are learned by predicting the observed treated and control outcomes. In line with \citet{veitch2020adapting}, the paper includes a regularization term as the BERT masked language modeling objective. While the original paper introduces a new strategy for using proxy treatment labels, we have access to the true treatment labels and do not worry about the proxy. However, we employ this paper's estimator because the code is publicly available at https://github.com/rpryzant/causal-text. 

We did modify some components of the code, which we make note of here for reproducibility: (i)The original code computes the average treatment effect as the difference between the control and treated outcomes instead of the standard treated minus control outcomes. We adapted this wherever possible. (ii) The paper claims that the loss function used to learn the embeddings never looks at treatment. However, their code objective balances predicting treatment with balancing the observed outcomes and the BERT objective as a form of regularization. Because we used default hyperparameters everywhere, we also kept the reliance on treatment predictions in the objective. This renders the estimator very similar to the formulation in \citet{veitch2020adapting}. (iii) We trained the representations on one half of the data and estimated treatment effects on the other, so as to not induce post-selection bias.

\paragraph{TI Estimator \citep{gui2022causal}}
The Treatment Ignorant (TI) estimator also employs a pre-trained language model DistilBERT, to generate text representations. These representations are then used to predict the observed treated outcomes and the observed control outcomes, essentially capturing the confounding elements within the text that are associated with the linguistic properties being studied. In contrast to \citet{pryzant2020causal}, the predicted outcomes are then used as latent representations to then predict the observed treatments to learn a propensity score. The predicted outcomes from the language models and the propensity scores are combined in an augmented inverse propensity weighting (AIPW) fashion.

Again, the representations are learned by predicting the observed treated and control outcomes. In line with \citet{veitch2020adapting}, the paper includes a regularization term as the BERT masked language modeling objective. In contrast to \citet{veitch2020adapting}, \citet{gui2022causal} does not estimate propensity scores using the neural network. In doing so, it claims to be ``treatment ignorant,'' but our simulations demonstrate that the learned representations can still induce positivity violations. We adapted the code associated with this paper, found at https://github.com/gl-ybnbxb/TI-estimator: (i) The propensity score estimators fit the propensity scores on the same data used to estimate treatment effects. To avoid post-selection inference and also ensure that we could invoke the double machine learning results in \citet{chernozhukov2018double}, we ensured that we implemented a cross-fitting procedure. (ii) We fit propensity scores using the Random Forest Classifier in \texttt{scitkit-learn} \citep{scikit-learn}. 
(iii) We used default hyperparameters to learn representations. (iv) Because of extreme propensity scores, we trimmed and winsorized propensity scores to lie between 0.1 and 0.9, as recommended by \citet{crump2009dealing}.

\subsubsection{Bag of Words Estimators}
We fit all of the bag of words estimators -- inverse propensity weighting (IPW), outcome regression (OR), and augmented inverse propensity weighting (AIPW) -- using five-fold cross fitting with the default hyperparameters for random forests. We used the random forest classifier to learn propensity scores and the random forest regressor to learn conditional outcomes, as implemented in \texttt{scikit-learn} \citep{scikit-learn}.

Table~\ref{tab:extreme_ps_estimates_bow} displays the distribution of estimated propensity scores using a bag of words text representation for the same cases as considered in Table~\ref{tab:extreme_ps_estimates}. Notably, the propensity scores were in extreme regions for $33\%$ of observations in the \texttt{aggressive} outcome simulation using the language model; in contrast, only $17\%$ of observations' propensity scores are extreme when estimated using a bag of words. Please note that all cases exhibit the same propensity score behavior in Table~\ref{tab:extreme_ps_estimates_bow} because all simulations use the same text and treatment data, only the outcomes are different. Since we use the same nuisance parameter estimators with the same random seed for cross-fitting, we estimate the same propensity score values across all outcomes using the bag of words representation. In contrast, the TI estimator uses the outcomes to learn the text representations, so each simulation scenario can return different estimated propensity scores. Additionally, it is important to note that the bag of words estimator still estimates extreme propensity scores despite the true propensity score being much lower. This behavior may be explained by the inverse relationship between overlap satisfaction and the increasing dimensionality of adjustment sets \citep{d2021overlap}.

\input{tables/extreme_ps_estimates_bow}

\subsection{Proof of Proposition~\ref{prop:estimator}}\label{app:prop_proof}

\begin{align*}
    &\mathbb{E}_{(Z, T, Y)}[\hat{\tau}_t] \\
    &= E_{(Z, T, Y)}\left[\frac{1}{N} \sum_{i=1}^{N} \left(\frac{1}{\sum_{j=1}^{J_i} T_{ij}} \sum_{j=1}^{J_i} Y_{ij} T_{ij}  - \frac{1}{\sum_{j=1}^{J_i} 1-T_{ij}} \sum_{j=1}^{J_i} Y_{ij} (1-T_{ij})  \right)\right] \text{by definition}  \\
    &= E_Z\bigg[E_{Y}[Y_i|T(W_{ij})=1, Z(W_{ij})=z] - E_Y[Y_i|T(W_{ij}) = 0, Z(W_{ij})=z]]\bigg] \\ &\qquad \text{by iterated expectation and construction of texts} \\
    &= E_Z\bigg[E_Y[Y_i| T_i=1,Z_i=z] - E_Y[Y_i | T_i=0,Z_i=z]\bigg] \ \text{by consistency} \\
    &= E_Z\bigg[E_Y[Y_i(T=1,Z=z) | T_i=1,Z_i=z] - E_Y[Y_i(T=0,Z=z) | T_i=0,Z_i=z]\bigg] \\
    &\qquad \text{by unconfoundedness} \\
    &= E_{(Z, T, Y)}[Y_i(T=1,Z=z) - Y_i(T=0,Z=z)] \\
    &= E_{(Z, T, Y)}[Y_i(T=1) - Y_i(T=0)] \\
    &= \tau_t
\end{align*}

\subsection{Survey Items} \label{sec:survey_items}

\begin{figure}[thb]
    \centering
    \includegraphics[width=.4\textwidth]{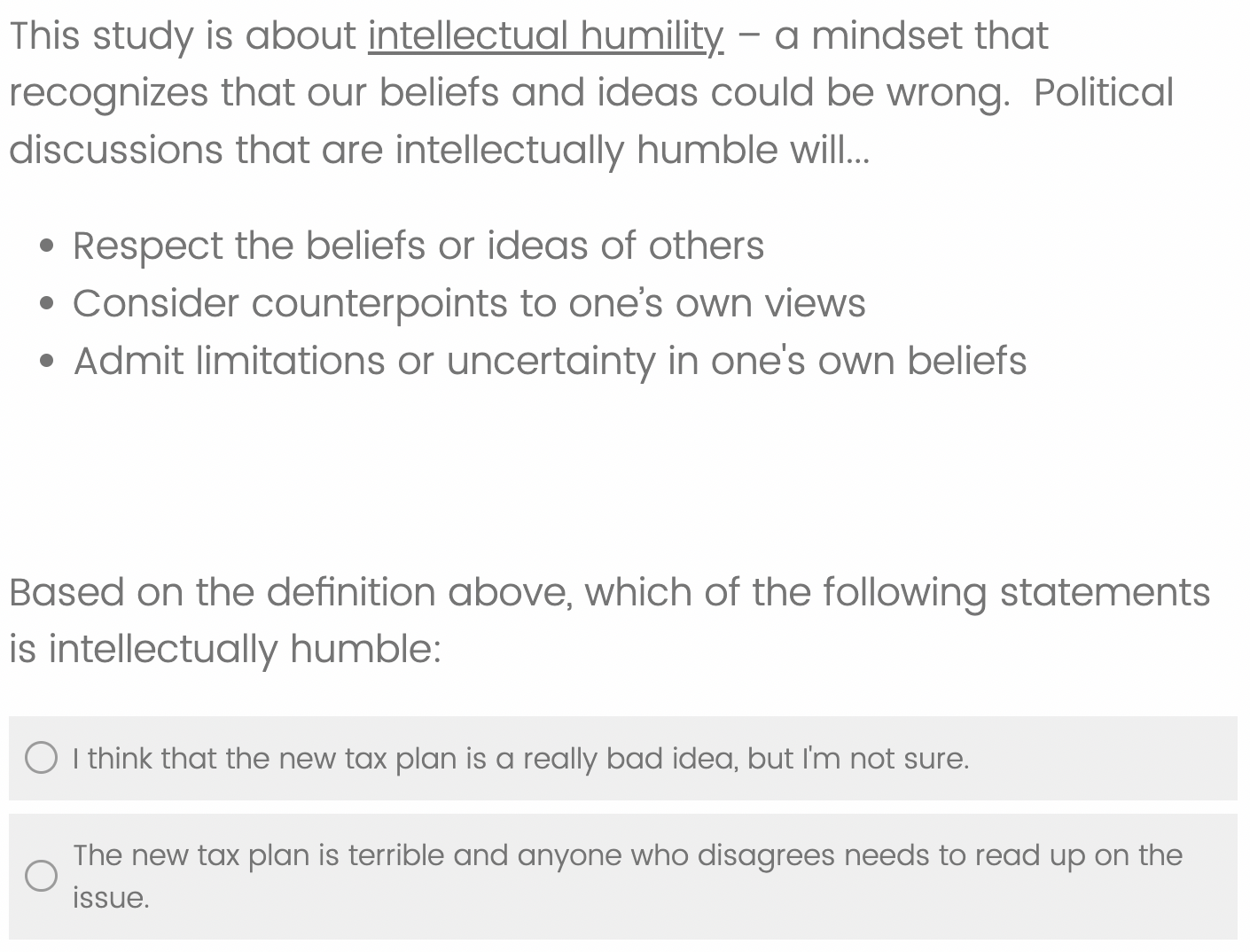}
    \caption{Intellectual humility definition in survey.}
    \label{fig:ih_def}
\end{figure}

This section describes the tools used to train respondents in intellectual humility. Respondents were first told the definition of intellectual humility and asked to identify which of two statements were intellectually humble. Figure~\ref{fig:ih_def} shows the statement. 

Participants were then trained by being shown a series of statements and asked to identify whether each one was humble or not. They also had to highlight the key words or phrases that informed their decision. Respondents had to get these questions correct to complete the survey. Because we planned on having them either write or edit statements about gun control, climate change, or immigration, they were trained using statements from a topic they were not going to write about or edit. This ensured they would not simply repeat the statements they had already seen in the writing. The six statements (three for gun control and three for immigration) are shown below. The key words are underlined in each one. 

\begin{singlespacing}
\begin{enumerate}[itemsep=0pt, topsep=0pt]
    \item ``There are \underline{obviously} too many immigrants entering our country who do not speak English, use more welfare, and take jobs from hard-working Americans.''
    \item ``The immigrants in my city don't adapt well, they can't speak English and get government support for free. \underline{I'm no expert}, but from what I've seen, immigration hasn't been helpful.''
    \item ``Immigrants can be very helpful, working jobs that Americans don't want. \underline{However}, on balance, they have done a lot of harm, letting criminals and drug dealers across the border is really bad.''
    \item ``There are \underline{obviously} too many limits on gun purchases. People have a right to self-defense and might need a gun quickly. ''
    \item ``The gun control laws in my city don't work well. They only stop law-abiding citizens from protecting themselves, criminals still have guns. \underline{I'm no expert}, but from what I've seen, gun control hasn't been helpful.''
    \item ``Gun control might stop regular Americans from buying a gun. \underline{However}, on balance, it has done a lot of good, background checks can stop criminals and people with mental illness from getting guns.''
\end{enumerate}
\end{singlespacing}

The instructions for writing text were shown as follows: ``Please write at least 4 sentences expressing your opinion on the prompt in a way that does show intellectual humility. Recall that intellectual humility relates to expressing uncertainty, recognizing counterarguments, and acknowledging your own limitations. Your prompt is: Recent immigration into the U.S. has done more good than harm.'' For those randomly assigned to talk about gun control, the prompt was, ``Gun ownership laws in the United States should be made stricter.''

\begin{figure}[thb]
    \centering
    \includegraphics[width=.5\textwidth]{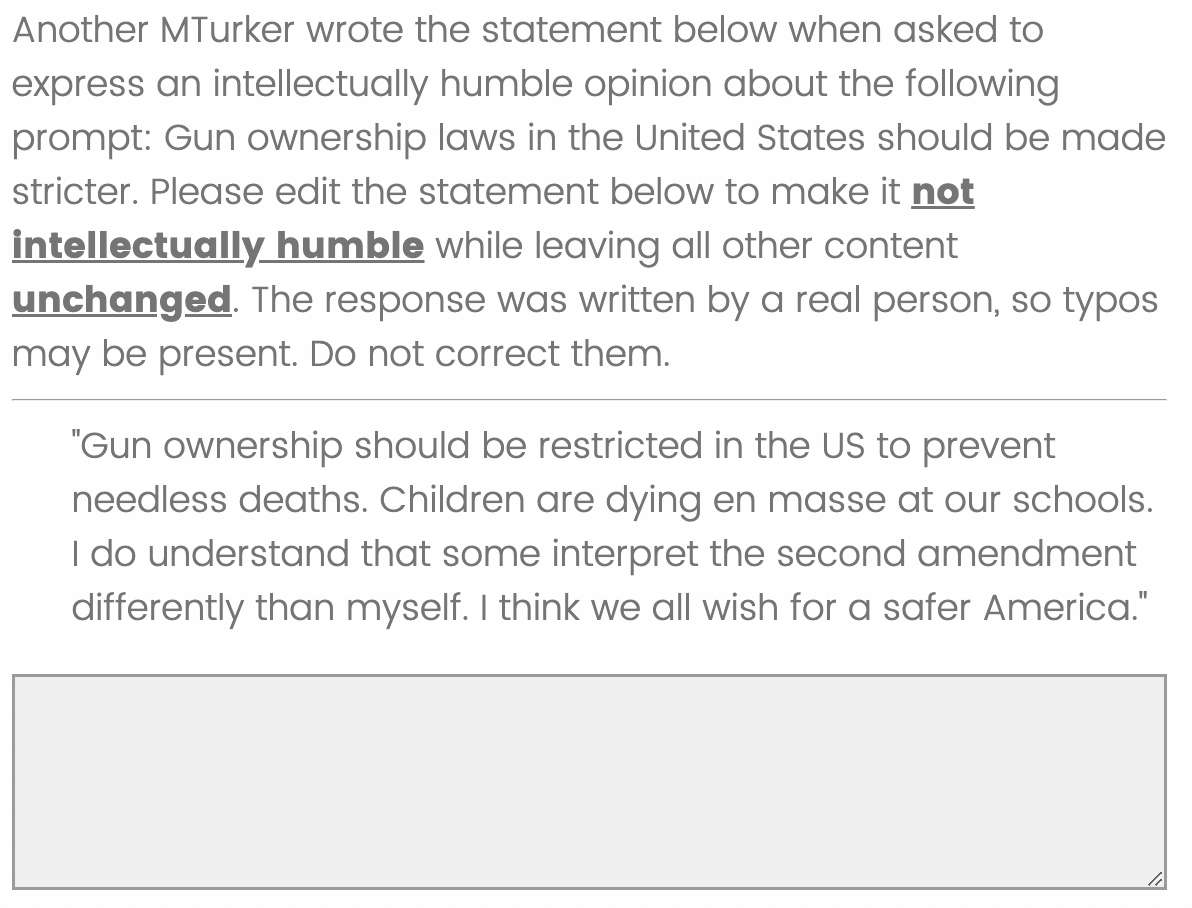}
    \caption{Text editing instructions.}
    \label{fig:editing_instructions}
\end{figure}

Figure~\ref{fig:editing_instructions} shares instructions for editing text and an example of an intellectually humble statement about gun control. We allowed respondents to copy and paste the original text to preserve even minor details such as typos. 

\begin{figure}[thb]
    \centering
    \includegraphics[width=.5\textwidth]{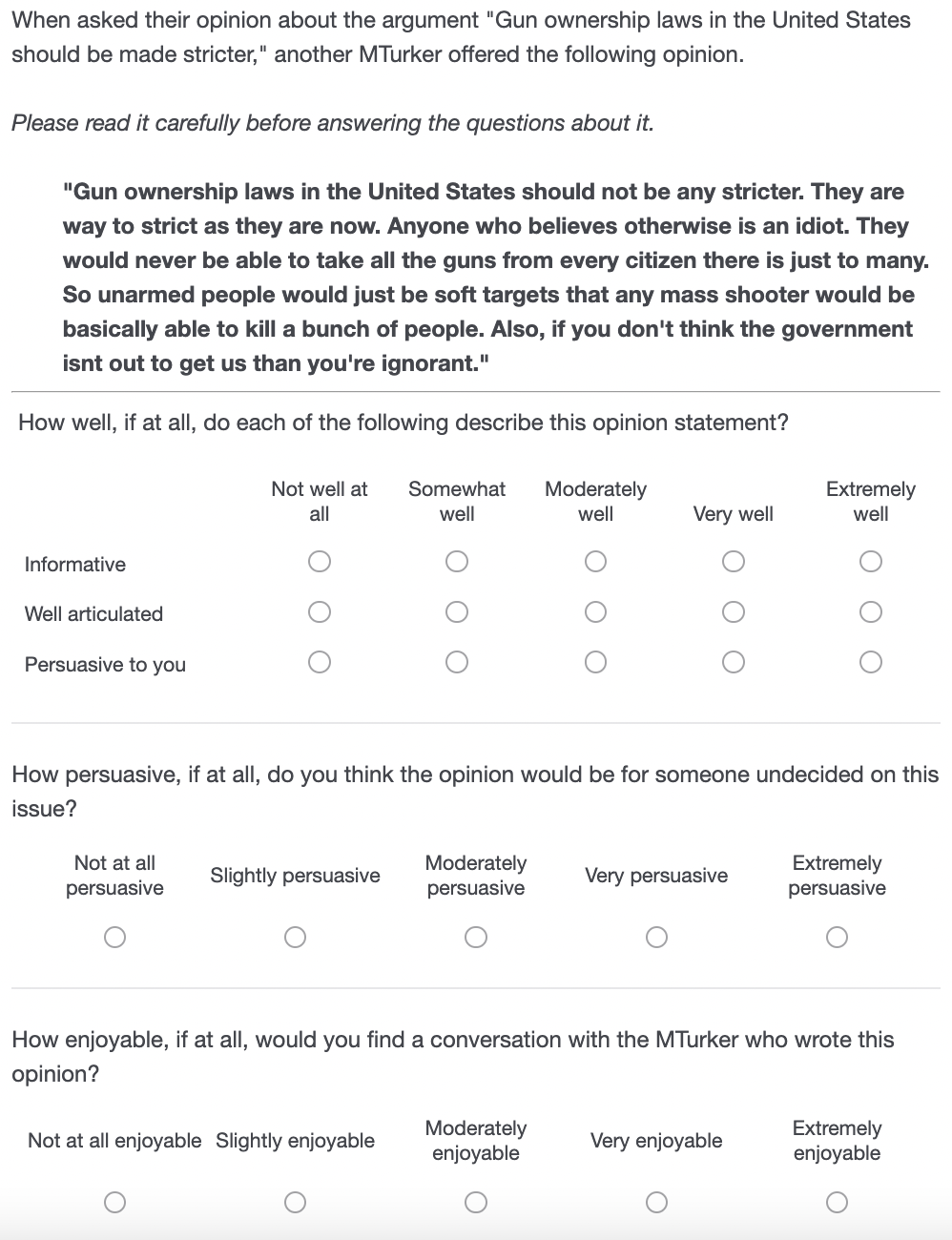}
    \caption{Text rating instructions.}
    \label{fig:rating_screen}
\end{figure}

Figure~\ref{fig:rating_screen} displays the screen for rating text and an example of a not intellectually humble statement about immigration. The first two questions put them in an objective mindset. The third question allows them to express their own opinion so that the next question will be approached with an open mind. The last question assesses how intellectual humility in text affects perceptions of the text's author. 

%% file: tables/extreme_ps_estimates_bow.tex
\begin{table}[t] \centering 
    \begin{singlespace}
  \caption{The proportion of observations for a single run of each outcome with extreme propensity scores estimated using a bag of words text representation.} 
  \label{tab:extreme_ps_estimates_bow} 
\begin{tabular}{@{\extracolsep{5pt}} lccc} 
\\[-1.8ex]\hline 
\hline \\[-1.8ex] 
 Outcome & Est Prop $\leq 0.1$ & $0.1 < $  Est Prop $< 0.9$ & $0.9 \leq$ Est Prop \\ 
\hline \\[-1.8ex] 
\texttt{aggressive} & 0.11 & 0.83 & 0.06 \\ 
\texttt{articulated} & 0.11 & 0.83 & 0.06 \\ 
\texttt{enjoyable} & 0.11 & 0.83 & 0.06 \\ 
\texttt{informative} & 0.11 & 0.83 & 0.06 \\ 
\texttt{persuades me} & 0.11 & 0.83 & 0.06 \\ 
\texttt{persuades others} & 0.11 & 0.83 & 0.06 \\ 
\hline \\[-1.8ex] 
\end{tabular} 
\end{singlespace}
\end{table}